\documentclass[preprint,aps,groupedaddress,showpacs]{revtex4-1}
\usepackage{amsmath,amssymb,bm,graphicx,epsfig,psfrag,ulem,color,natbib}

\newcommand{\un}{\mathbf {u}^n}
\newcommand{\us}{\mathbf {u}^s}
\newcommand{\nus}{\nu^s}
\newcommand{\nun}{\nu^n}
\newcommand{\uu}{\mathbf {u}}
\newcommand{\Fns}{\mathbf {F}}
\newcommand{\rhos}{\rho^s}
\newcommand{\rhon}{\rho^n}

\newcommand{\etal}{{\it et al.}~}
\newcommand{\bom}{{\mbox{\boldmath $\omega$}}}
\newcommand{\boms}{{\mbox{\boldmath $\omega$}}^s}

\begin{document}

\title{Shell model of superfluid turbulence}

\author{D.~H.~Wacks}
\email{d.h.wacks@ncl.ac.uk}
\author{C.~F.~Barenghi} 
\affiliation{School of Mathematics and Statistics, University of
Newcastle, Newcastle upon Tyne, NE1 7RU, UK}


\begin{abstract}
Superfluid helium consists of two inter-penetrating fluids, a viscous normal fluid and an inviscid superfluid, coupled by a mutual friction. We develop a two-fluid shell model to study superfluid turbulence. We investigate the energy spectra and the balance of fluxes between the two fluids as a function of temperature in continuously forced turbulence, and, in the absence of forcing, the decay of turbulence. We furthermore investigate deviations from the $k^{-5/3}$ spectrum caused by the mutual friction force. We compare our results with experiments and existing calculations. We find that, at sufficiently low temperatures a build-up of energy develops at high wavenumbers suggesting the need for a further dissipative effect, such as the Kelvin wave cascade and phonon emission.
\end{abstract}


\maketitle
\section{Introduction}

Helium~II (the low temperature, quantum phase of liquid $^4$He) consists
of two inter-penetrating fluid components \cite{Donnelly}: 
the inviscid superfluid (associated
with the quantum ground state) and the viscous normal fluid (consisting of
thermal excitations). 
Each fluid component has its own density and velocity
field, $\rhos$, $\us$ for the superfluid and $\rhon$, $\un$ for the
normal fluid, where $\rho=\rhos+\rhon$ is helium's density and
the relative proportion of superfluid and normal fluid depends on
the absolute temperature $T$.

What makes helium~II particularly interesting is that the 
superfluid vorticity is concentrated in thin, discrete vortex filaments
of fixed (quantized) circulation 
$\kappa=h/m=9.97 \times 10^{-4}~\rm cm^2/s$, where
$h$ is Planck's constant and $m$ the mass of one $^4$He atom. 
The vortex filaments 
interact with the thermal excitations, 
causing a mutual friction between superfluid and normal fluid \cite{BDV}.

Turbulence in helium~II (quantum turbulence) is easily generated
in the laboratory by stirring or pushing liquid helium with 
grids and propellers \cite{Oregon,Roche,Salort}, vibrating
forks \cite{Prague} or other means. 
Quantum turbulence is a complex state \cite{Vinen-Niemela} in which 
normal fluid eddies of arbitrary shapes and strengths 
interact with a tangle of discrete vortex filaments of fixed strength. 
Despite the two-fluids nature of quantum turbulence, experiments have
shown remarkable similarities with turbulence in ordinary fluids
(classical turbulence), for example the same pressure drops
along pipes and channels \cite{Walstrom}, drag crisis behind 
a sphere \cite{drag-crisis}, and Kolmogorov energy spectrum \cite{Tabeling}.
The last property, which is our particular concern in this paper, describes how the energy of the flow is distributed over the length scales in homogeneous isotropic turbulence.

Whereas turbulence in ordinary fluids is
studied on the firm ground of the Navier-Stokes equation, there is not
such an established equation for turbulent helium~II.
Existing numerical simulations \cite{Bag11a,Adachi}
of quantum turbulence follow the approach
of Schwarz \cite{Schwarz}, who calculated the properties of the vortex tangle 
by integrating in time the motion of a great number of
individual vortex filaments in the presence of a prescribed normal 
fluid, neglecting the back reaction of the superfluid onto
the normal fluid. 

The aim of this work is to gain insight into self-consistent
turbulent two-fluids hydrodynamics using a shell model \cite{Biferale}. 
In classical fluid dynamics \cite{Frisch} shell models are idealized truncated models 
of the Navier-Stokes equation which
neglect any geometry of the flow but describe the Richardson
cascade, its properties (e.g. the Kolmogorov
energy spectrum) over a wide range of scales and times.
In the less understood context of helium~II, shell models 
allow us to explore the interaction of normal fluid and superfluid 
in the inertial and dissipative ranges in a relatively simple way.
What happens at very large length scales \cite{adzhemyan} where
finite-size effects may affect the dynamics of the energy-containing
eddies is beyond the scope of this work.

\section{Model}
\subsection{Classical GOY model}

The classical Navier-Stokes equation is

\begin{equation}
\rho (\frac{\partial \uu}{\partial t}+\uu \cdot \nabla \uu)=
- \nabla p  
+\mu \nabla^2 \uu, 
\label{eq:NS}
\end{equation}

\noindent
with the incompressibility condition $\nabla \cdot \uu=0$, where
$\uu$ is the velocity, $p$ the
pressure, $\rho$ the density and $\mu$ the viscosity.
Starting from the works of 
Gledzer \cite{Gledzer} and Yamada \& Okhitani \cite{Yamada-Okhitani},
a variety of shell models have been developed \cite{Sabra}
to study turbulent solutions of Eq.~\ref{eq:NS}.
The GOY model, named after the pioneering authors, 
is a finite-dimensional dynamical system 
which describes the dynamics of idealised (complex)
Fourier components of the velocity field, $u_m$ corresponding
to wavenumber $k_m$; the index $m$ ($m=1,\cdots,M$) is
called the shell index.
The wavenumbers are geometrically
distributed in k-space, $k_m=k_0 \lambda^m$, where $k_0$ is a
reference wavenumber and $\lambda>1$. The equations of
motion are

\begin{equation}
\left( \frac{d}{dt}+\nu k_m^2 \right) u_m 
=G_m[u] + f \delta_{m,m'},
\label{eq:goy}
\end{equation}

\noindent
for $m=1,\cdots M$,
where $\nu=\mu/\rho$ is the kinematic viscosity, 
$\delta_{m,m'}$ is Kronecker's delta, and $f$ is the 
amplitude of external forcing applied on a particular shell $m=m'$.
The inertial term $G_m[u]$ is quadratically
nonlinear and local in k-space, coupling $u_m$ with its nearest
neighboring shells;
more precisely we have

\begin{equation}
G_m[u]=i(c_m^{(1)} \bar{u}_{m+1} \bar{u}_{m+2}
    +c_m^{(2)} \bar{u}_{m-1} \bar{u}_{m+1}
    +c_m^{(3)} \bar{u}_{m-1} \bar{u}_{m-2}),
\label{eq:G}
\end{equation}

\noindent 
where $\bar{u}$ denotes the complex conjugate of $u$, and
\begin{equation}
c_m^{(1)}=ak_m, \qquad c_m^{(2)}=b k_{m-1}, \qquad c_m^{(3)}=c k_{m-2},
\label{eq:3}
\end{equation}

\noindent
The boundary conditions are $u_m=0$ for $m\leq0$, $m>M$
and
\begin{equation}
c_1^{(2)}=c_1^{(3)}=c_2^{(3)}=c_{M-1}^{(1)}=c_M^{(1)}=c_M^{(2)}=0,
\end{equation}

The coefficients $a=1$, $b=-1/2$,  $c=-1/2$ and $\lambda=2$
are chosen so that in the steady ($d/dt=0$), 
unforced ($f=0$), inviscid ($\nu=0$) case the nonlinear interaction
conserves the two quadratic invariants of the 3-dimensional Euler equation,
energy and helicity,
\begin{equation}
E=\frac{1}{V} \int \frac{1}{2} \uu \cdot \uu dV,
\qquad
H=\frac{1}{V} \int \bom \cdot \uu dV,
\end{equation}

\noindent
(where $\bom=\nabla \times \uu$ and $V$ is volume) which are identified as

\begin{equation}
E=\frac{1}{2} \sum_{m=1}^M \vert u_m \vert^2
 =\sum_{m=1}^M E_m k_m,
\label{eq:energy}
\end{equation}

\noindent
where

\begin{equation}
E_m=\frac{\vert u_m \vert^2}{2 k_m},
\end{equation}
is the spectral energy associated
with shell $m$, and

\begin{equation}
H=\frac{1}{2} \sum_{m=1}^M (-1)^m \lambda^m  \vert u_m \vert^2.
\label{eq:helicity}
\end{equation}

\subsection{Two-fluids equations}

The two-fluids equations of helium~II are \cite{Donnelly}

\begin{equation}
\rhos (\frac{\partial \us}{\partial t}+\us \cdot \nabla \us)=
-\frac{\rhos}{\rho} \nabla p +\rhos S \nabla T -\Fns
\label{eq:us}
\end{equation}

\begin{equation}
\rhon (\frac{\partial \un}{\partial t}+\un \cdot \nabla \un)=
-\frac{\rhon}{\rho} \nabla p -\rhos S \nabla T 
+\mu \nabla^2 \un +\Fns
\label{eq:un}
\end{equation}

\noindent
with $\nabla \cdot \us=\nabla \cdot \un=0$,
where  $S$ is the specific entropy. Hereafter we denote by  $\nun=\mu/\rhon$
the kinematic viscosity of helium~II.
If $\Fns=0$ and $\nabla \times \us=\boms=0$,
Eqs.~\ref{eq:us} and \ref{eq:un} reduce to Landau's two-fluids equations,
which describe the well-known mechanical and thermal behaviour of 
helium~II in the absence of vortex lines. In the
presence of vortices, the mutual friction $\Fns$ which couples the two fluids
depends on the vortex line density $L$ (vortex length per unit volume), or, more precisely, on some suitably defined coarse-grained vorticity field $\boms$. Unfortunately
the form of $\Fns$ is uncertain. The form given by Hall \& Vinen \cite{Hall}
accounts for experiments in rotating cylinders and Taylor-Couette flow
 \cite{Barenghi-Couette}, 
in which vortices are either straight or curved, but still
polarised in the same direction. In the case of turbulent flow, vortex filaments may be random or partially polarised, so the relation between the coarse-grained vorticity $\boms$ and the vortex line density $L$ is not clear. Hereafter for the sake
of simplicity, we assume the Gorter-Mellink \cite{gorter-mellink} form

\begin{equation}
\Fns=\frac{B \rhos \rhon}{2 \rho} \kappa L (\us -\un),
\label{eq:friction}
\end{equation}

\noindent
where $B$ is a known temperature dependent mutual friction coefficient.
This form for the mutual friction is motivated by dimensional arguments and is consistent with the method employed to measure vortex line
density in experiments. It has been widely used in the quantum turbulence literature \cite{tough,BDV,melotte,Vinen-Niemela,vinen3he,Leveque}.

\subsection{Two-fluids GOY model}

The natural generalization of Eq.~\ref{eq:goy} to two fluids is

\begin{equation}
\frac{d}{dt} u_m^s 
=G_m[u^s] - F_m + f^s \delta_{m,m'},
\label{eq:goy-us}
\end{equation}

\begin{equation}
\left( \frac{d}{dt}+\nun k_m^2 \right) u_m^n 
=G_m[u^n] + \frac{\rhos}{\rhon}F_m + f^n \delta_{m,m'},
\label{eq:goy-un}
\end{equation}

\noindent
where the nonlinear term for each fluid is as in Eq.~\ref{eq:G},
and the mutual friction is
\begin{equation}
F_m=\alpha \kappa L(u_m^s-u_m^n).
\end{equation}

\noindent
The temperature dependence of the friction coefficient
$\alpha=B\rhon/(2 \rho)$ is well-known \cite{Donnelly-Barenghi}.
For consistency, the vortex line density $L$ is not an arbitrary
parameter, but is identified as
$L=Q^{1/2}/\kappa$ where the superfluid enstrophy
$Q$ is 
\begin{equation}
Q=\sum_{m=1}^M \frac{1}{2} k_m^2 \vert u_m^s \vert^2.
\end{equation}

Clearly this model, which describes superfluid vorticity as a 
continuum, is meaningful only for length scales bigger than the average
intervortex spacing $\ell \approx L^{-1/2}$, that is to say for
wavenumbers $k_m<k_{\ell}=1/{\ell}=L^{1/2}$.

In a recent paper, Roche \etal \cite{Leveque} solved
equations similar to our Eqs.~\ref{eq:us} and \ref{eq:un}
in the presence of continuous forcing applied to the fluid with the greater density, 
thus performing the first
Direct Numerical Simulations (DNS) of two-fluids hydrodynamics.
Unlike our two-fluids shell model, their two-fluids DNS
contained an artificial superfluid
viscous force $\nus \rhos \nabla^2 \us$ at the right hand side
of Eq.~\ref{eq:us}. The introduction of this unphysical term was motivated
by numerical analysis. They set the ratio $\nun/\nus=4$ 
to minimize the effect of the artificial viscosity while preserving
the stability of the calculation. We stress that our two-fluids shell
model (see Eq.~\ref{eq:goy-us}) does not need such an artificial term.
 Another related model is the large-eddy calculation of Merahi \etal \cite{Merahi}

\section{Results}
\subsection{Forced Turbulence}
Unless stated otherwise, the results which we present are obtained
by numerically integrating in time Eqs.~\ref{eq:goy-us} and \ref{eq:goy-un}
for $M=18$, $k_0=2^{-4}$. The same forcing 
(typically $f^s=f^n=(1+i) \times 5 \times 10^{-3}$) is applied to shell $m=4$
for both superfluid and normal fluid. A typical initial condition is
$u_m^j=(1+i) k_m \exp{(-k_m^2/2)}$,$~(j=n,s)$, for which $E_m^j=k_m \exp{(-k_m^2)}$.
The time stepping combines the Crank-Nicolson method
for the diffusion term and the Adams-Bashforth method for
the other terms; the typical time step is $\Delta t = 5 \times 10^{-6}$.
We stop the time integration after the spectrum saturates, making sure
that the elapsed time is of the order of 10 large eddy turnover times
(about 100 turnover times of the forcing wavenumber), which is
typically of the order of $500~\rm s$. We find that, once
saturation is achieved, time averaged spectra are the same as ensemble
averaged spectra (obtained using randomly phase shifted realisations).

To study the temperature dependence of the results, we consider
three temperatures $T=2.157~\rm K$, $1.96~\rm K$ and $1.44~\rm K$
which hereafter we refer to as high, medium and low temperatures \cite{Leveque}.
The values of $\rho^s$, $\rho^n$, $\rho$, $\nu$ and $\alpha$ in TABLE 1. corresponding to these temperatures are
\begin{table}[h]
\begin{center}
\begin{tabular}{|l|c|c|c|c|c|} \hline
$T$ (K) & $\rho^s$ (g/cm$^3$) & $\rho^n$ (g/cm$^3$) & $\rho$ (g/cm$^3$)& $\nu$ (cm$^2$/s) & $\alpha$ \\ \hline
$2.157$ & $0.01510$ & $0.12939$ & $0.14449$ & $1.526\times10^{-4}$ & $1.045$\\
$1.96$ & $0.07335$ & $0.07221$ & $0.14556$ & $9.694\times10^{-5}$ & $0.245$\\
$1.44$ & $0.13251$ & $0.01264$ & $0.14515$ & $9.538\times10^{-5}$ & $0.059$\\ \hline
\end{tabular}
\caption{Helium parameters used in our calculations.}
\label{tab1}
\end{center}
\end{table}
The resulting normal fluid fractions, $\rho^n/\rho^s$, are approximately $10$, $1$ and $0.1$ for high, medium and low temperatures respectively. 

\begin{figure}
	\centering
	\includegraphics[width=0.6\textwidth]{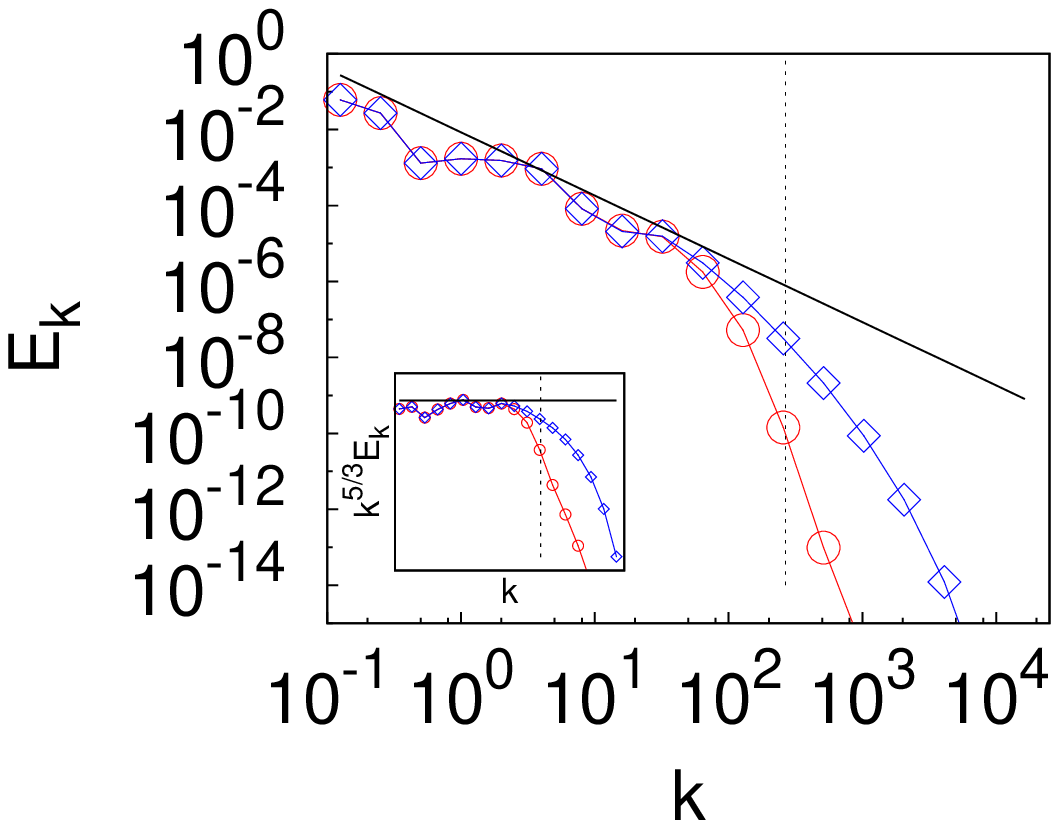}
	\includegraphics[width=0.6\textwidth]{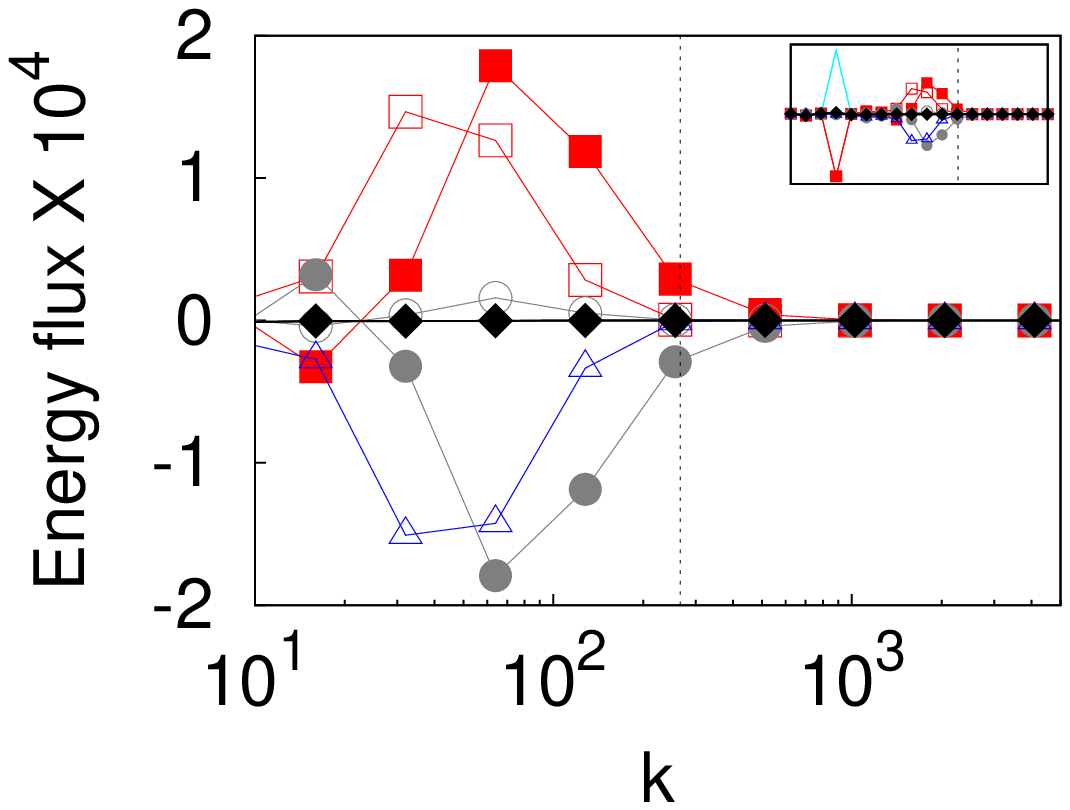}
	\caption{(Colour online).
	Top:
	Main plot: Log-log plot of superfluid (blue diamond) and normal fluid (red circle) energy spectra $E_k$ ($\rm{cm}^2~\rm{s}^{-2}$) vs wavenumber $k$ ($\rm{cm}^{-1}$) at $T=2.157~\rm{K}$. The solid line denotes the $k^{-5/3}$ Kolmogorov spectrum. The vertical dotted line marks $k_{\ell}$.
	Inset: Compensated spectra. As above, but $k^{5/3} E_k$ vs $k$.
	Bottom:
	Main plot: Log-lin plots of time-averaged scale-by-scale energy budget $dE_k/dt$ ($\rm{cm}^2~\rm{s}^{-3}$) vs wavenumber $k$ ($\rm{cm}^{-1}$) for the two-fluid model at high temperature ($2.157\rm{K}$). We show the fluxes for the normal fluid (hollow shapes) and superfluid (solid shapes). The main plot shows the balance of fluxes in the normal fluid between the inertial term $T^n_m$ (red squares) and the viscous term $D^n_m$ (blue triangles), and in the superfluid between the inertial term $T^s_m$ (red squares) and the mutual friction term $M^s_m$ (grey circles). We also show the total flux $dE_m/dt$ (black diamonds).
	Inset: In the inset we show the fluxes over the entire wavenumber range, including the flux due to the external forcing $\epsilon^{n,s}_{inj}$ (light blue line) which is nonzero only at shell $n=4$. The vertical dotted line denotes the time-averaged wavenumber, $k_{\ell}$, corresponding to the intervortex spacing.}
\label{fig1}
\end{figure}

Fig.~\ref{fig1} (top) shows superfluid and normal fluid spectra at
high temperature. It is apparent that superfluid and normal fluid
velocity lock onto each other by the mutual friction over many length scales as envisaged by Vinen and Niemela \cite{Vinen-Niemela} and Barenghi \etal \cite{BHS}. Furthermore a closer inspection of the complex $u_m$ throughout the inertial subrange reveals that $\Re\left(u^n_m\right)=\Re\left(u^s_m\right)$ and $\Im\left(u^n_m\right)=\Im\left(u^s_m\right)$ to at least $4$dp. Both spectra are consistent with the Kolmogorov
scaling $k^{-5/3}$ (denoted by the solid line) over a wide inertial range $k_0~\ll~k\ll~k_{\ell}$, as observed in experiments \cite{Tabeling} and DNS of two-fluids hydrodynamics \cite{Leveque}.
The vertical dotted line denotes the wavenumber 
$k_{\ell}$ which corresponds to the intervortex spacing. 
To make the Kolmogorov scaling more evident, we plot the compensated
spectra $k^{5/3}~E_k$ (see Fig.~\ref{fig1} (top,inset)).
It is apparent that the superfluid spectrum extends to bigger 
wavenumbers than the normal fluid's: this is because there are no 
viscous forces acting on the superfluid; nevertheless, the superfluid
spectrum decays at large $k$ because superfluid motion 
is damped by mutual friction. We confirmed that the development of a $k^{-5/3}$ spectrum is dependent neither on our choice of initial condition nor on the fact that we force both fluids.

It is instructive to consider the scale-by-scale energy budget per unit mass.
Using Eqs.~\ref{eq:goy-us} and \ref{eq:goy-un} and the fact that
\begin{equation*}
\frac{dE_m}{dt}=\frac{1}{2}\left(\frac {du_m}{dt}~\bar{u}_m+u_m~\frac{d\bar{u}_m}{dt} \right)
\end{equation*}
we obtain
\begin{eqnarray}
	\frac{dE^n_m}{dt} &=& T^n_m + D^n_m + M^n_m + \epsilon^n_{inj}\delta_{m,4},
	\\[12pt]
	\frac{dE^s_m}{dt} &=& T^s_m + M^s_m + \epsilon^s_{inj}\delta_{m,4},
\end{eqnarray}
\noindent
where
\begin{equation}
T^j_m = -\Im\left(G_m[u^j]u^j_m\right)
\end{equation}
are the energy transfer rates arising from the triadic interactions between Fourier modes within each fluid,
\begin{equation}
D^n_m = -2\nun~k^2_mE^n_m
\end{equation}
is the rate of viscous dissipation in the normal fluid,
\begin{equation}
M^n_m = \frac{\rhos}{\rhon}\Re\left(F_m\bar{u}^n_m\right)
\end{equation}
and
\begin{equation}
M^s_m = -\Re\left(F_m\bar{u}^s_m\right)
\end{equation}
are the rates of exchange of kinetic energy between the two fluids due to mutual friction ($M^n_m$ being the flow of energy from the normal fluid to the superfluid and $M^s_n$ the flow from the superfluid to the normal fluid) and 
\begin{equation}
\epsilon^j_{inj} = f^n\delta_{m,m'}\Re\left(u^j_m\right)
\end{equation}
are the rate of influx of energy due to the forcing terms, where $\Re$ and $\Im$ denote real and imaginary parts respectively.
The nonlinear triadic interaction $T^j_m$ is defined such that the energy flux at shell $m^*$ is given by
\begin{eqnarray}
\Pi^j_m &=& \sum_{m\leq~m^*}T^j_m\\
        &=& \Delta^j_{n+1}-(-b-1)\Delta^j_n,
\end{eqnarray}
where we define the correlator
\begin{equation}
\Delta^j_n = k_{n-1}\Im\left(u^j_{n-1}u^j_nu^j_{n+1}\right),
\end{equation}
all other terms cancelling due to the conservation of energy.

Fig.~\ref{fig1} (bottom) shows the energy balance in the inertial range.
It is apparent that the total energy flux in each shell is
zero (solid black diamond).
We find that the normal fluid's inertial term 
(hollow red squares) is balanced by the viscous term (blue hollow
triangles), as in ordinary turbulence. On the contrary,
the superfluid's inertial term (solid red squares)
is balanced by the mutual friction (solid grey circles), in agreement
with Roche \etal \cite{Leveque}. The inset of Fig.~\ref{fig1} (bottom) shows
the energy balance over the entire k-space, including the
contribution of the forcing on the shell $m=4$.
We observe that $M^n_m$ and $M^s_m$ are always of opposite signs, as expected from their definitions,
and, almost without exception, $M^n_m>0$ and $M^s_m<0$. Furthermore $D^n_m<0$ for all $m$.

\begin{figure}
	\centering
	\includegraphics[width=0.6\textwidth]{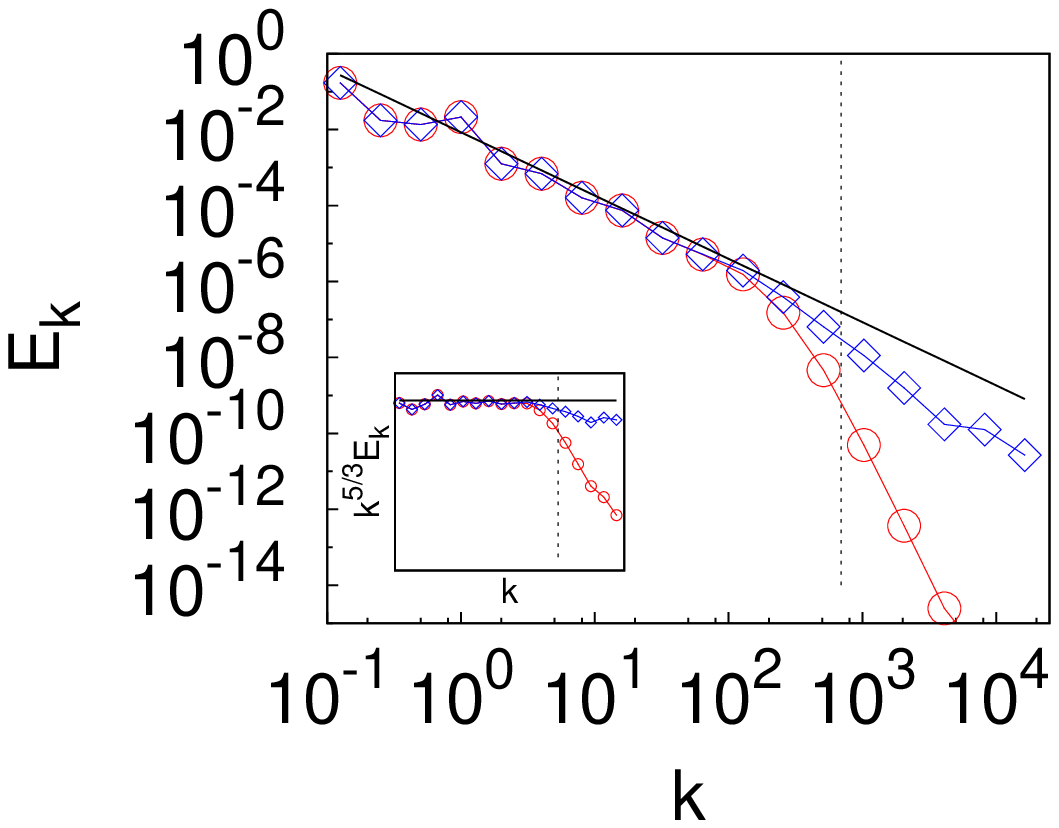}
	\includegraphics[width=0.6\textwidth]{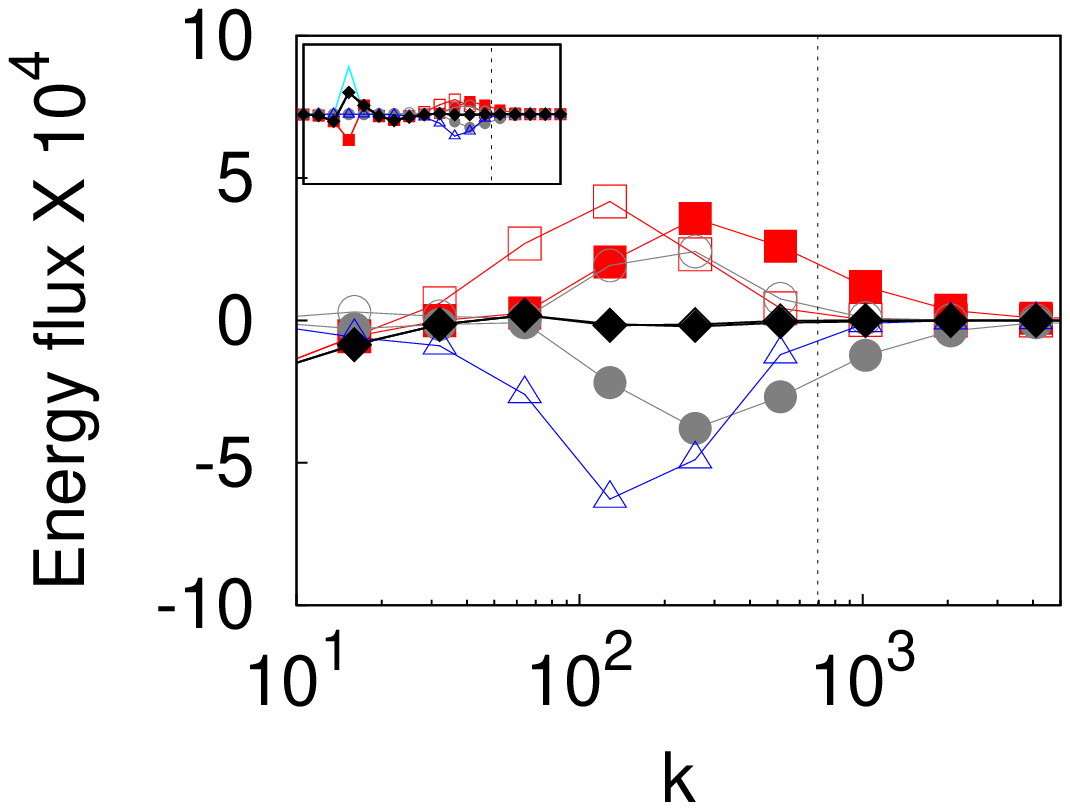}
	\caption{(Colour online).
	Top:
	Main plot: Log-log plot of superfluid (blue diamond) and normal fluid (red circle) energy spectra $E_k$ ($\rm{cm}^2~\rm{s}^{-2}$) vs wavenumber $k$ ($\rm{cm}^{-1}$) as in \ref{fig1} (top) but at $T=1.96~\rm{K}$. The solid line denotes the $k^{-5/3}$ Kolmogorov spectrum. The vertical dotted line marks $k_{\ell}$.
	Inset: Compensated spectra. As above, but $k^{5/3} E_k$ vs $k$.
	Bottom:
	Main plot: Log-lin plots of time-averaged scale-by-scale energy budget $dE_k/dt$ ($\rm{cm}^2~\rm{s}^{-3}$) vs wavenumber $k$ ($\rm{cm}^{-1}$) for the two-fluid model as in \ref{fig1} (bottom) but at medium temperature ($1.96\rm{K}$). We show the fluxes for the normal fluid (hollow shapes) and superfluid (solid shapes). The main plot shows the balance of fluxes in the normal fluid between the inertial term $T^n_m$ (red squares) and the viscous term $D^n_m$ (blue triangles), and in the superfluid between the inertial term $T^s_m$ (red squares) and the mutual friction term $M^s_m$ (grey circles). We also show the total flux $dE_m/dt$ (black diamonds).
	Inset: In the inset we show the fluxes over the entire wavenumber range, including the flux due to the external forcing $\epsilon^{n,s}_{inj}$ (light blue line) which is nonzero only at shell $n=4$. The vertical dotted line denotes the time-averaged wavenumber, $k_{\ell}$, corresponding to the intervortex spacing.}
\label{fig2}
\end{figure}

\begin{figure}
	\centering
	\includegraphics[width=0.6\textwidth]{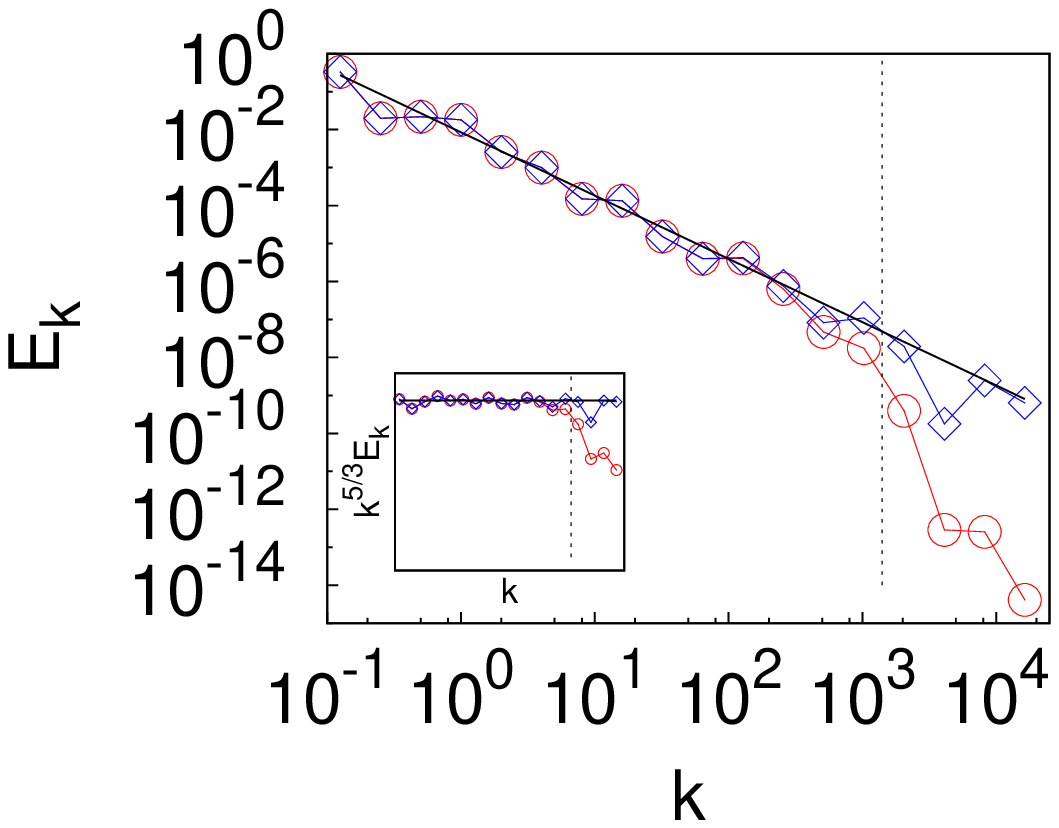}
	\includegraphics[width=0.6\textwidth]{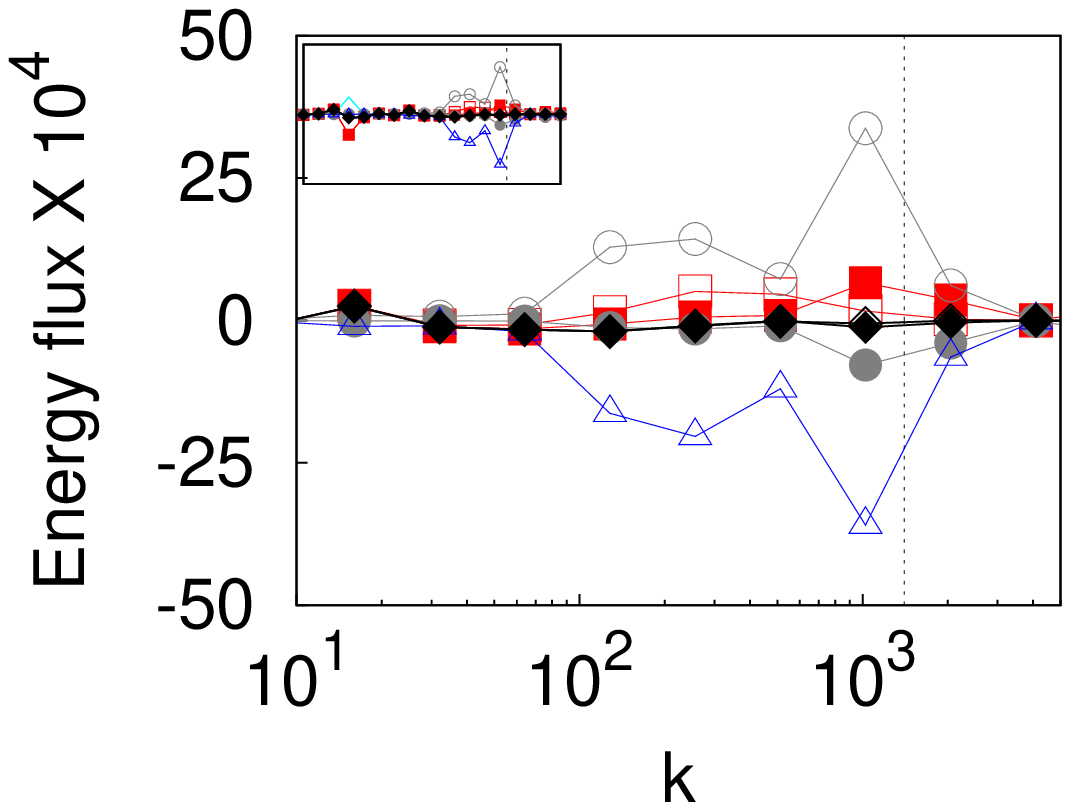}
	\caption{(Colour online).
	Top:
	Main plot: Log-log plot of superfluid (blue diamond) and normal fluid (red circle) energy spectra $E_k$ ($\rm{cm}^2~\rm{s}^{-2}$) vs wavenumber $k$ ($\rm{cm}^{-1}$) as in \ref{fig1} (top) but at $T=1.44~\rm{K}$. The solid line denotes the $k^{-5/3}$ Kolmogorov spectrum. The vertical dotted line marks $k_{\ell}$.
	Inset: Compensated spectra. As above, but $k^{5/3} E_k$ vs $k$.
	Bottom:
	Main plot: Log-lin plots of time-averaged scale-by-scale energy budget $dE_k/dt$ ($\rm{cm}^2~\rm{s}^{-3}$) vs wavenumber $k$ ($\rm{cm}^{-1}$) for the two-fluid model as in \ref{fig1} (bottom) but at low temperature ($1.44\rm{K}$). We show the fluxes for the normal fluid (hollow shapes) and superfluid (solid shapes). The main plot shows the balance of fluxes in the normal fluid between the mutual friction term $M^n_m$ (red squares) and the viscous term $D^n_m$ (blue triangles), and in the superfluid between the inertial term $T^s_m$ (red squares) and the mutual friction term $M^s_m$ (grey circles). We also show the total flux $dE_m/dt$ (black diamonds).
	Inset: In the inset we show the fluxes over the entire wavenumber range, including the flux due to the external forcing $\epsilon^{n,s}_{inj}$ (light blue line) which is nonzero only at shell $n=4$. The vertical dotted line denotes the time-averaged wavenumber, $k_{\ell}$, corresponding to the intervortex spacing.}
\label{fig3}
\end{figure}

The basic picture did not change when we reduced the temperature
to $T=1.96~\rm K$ (Fig.~\ref{fig2}), and then
to $T=1.44~\rm K$ (Fig.~\ref{fig3}), again in agreement with experiments \cite{Tabeling}.
There are however noticeable differences which develop as the temperature is lowered.
At high temperature viscous forces provide a sufficient energy sink for both normal fluid and superfluid. As the temperature drops the normal fluid fraction decreases and the superfluid fraction increases. This reduces the ability of the normal fluid viscosity to adequately dissipate the energy. As $T$ is decreased, both the normal fluid and superfluid spectra extend to the right, towards higher wavenumbers, and the wavenumber, $k_{\ell}$, corresponding to the intervortex spacing, moves progressively to the right. Furthermore at low $T$ in the normal fluid the viscous dissipation is no longer balanced by the inertial term, but by the mutual friction term (see Fig.~\ref{fig3} bottom).

In their pure states ($\rhos=0$ and $\rhon=0$ respectively) we expect the spectrum of the normal fluid to decay exponentially after $k_{\eta}$ (see Eq.~\ref{eq:k1}), the Kolmogorov microscale at which the inertial term is balanced by the viscous dissipation, and the superfluid to continue to cascade like $k^{-5/3}$; a build-up of energy at high $k$ is thus expected, although the physical interpretation of our model is limited to $k<k_{\ell}$. We can furthermore expect that as we approach these limiting cases the dominant fluid will cause a deviation in the spectrum of the other fluid. L'vov \etal \cite{lvov2006} derived approximate expressions for each of these cases.

At low temperatures ($\rhos\gg\rhon$), the normal fluid spectrum, instead of decaying exponentially for $k>k_{\eta}$, deviates slightly from $k^{-5/3}$ due to the force exerted on it by the mutual friction. This continues until the length scale, denoted $k_{\star}$ (see Eq.~\ref{eq:k2}), at which the viscous dissipation balances the mutual friction. Beyond this wavenumber the normal fluid spectrum decays with a power-law $k^{-17/3}$; not exponentially as in the case of classical turbulence. This power-law decay is valid at least until $k_{\ell}$, the wavenumber corresponding to the inter-vortex spacing, at which point the model for the superfluid is no longer valid.

On the other hand, at high temperatures ($\rhon\gg\rhos$), the superfluid spectrum beyond $k_{\eta}$ is affected by the exponentially decaying normal fluid so that it deviates from its $k^{-5/3}$ power-law to a steeper slope $k^{-3}$. This continues for as long as the mutual friction dominates over the superfluid inertial term. At some wavenumber, $k_+$ (see Eq.~\ref{eq:k3}), this relationship shifts such that the inertial term becomes dominant and the superfluid spectrum regains the $k^{-5/3}$ power-law for as long as the model remains valid ($k<k_{\ell}$).

Following L'vov \cite{lvov2006}, we define
\begin{eqnarray}
	k_{\eta} &=& \epsilon^{1/4}\nu^{-3/4},
	\label{eq:k1}
	\\[12pt]
	k_{\star} &=& \alpha^{1/2}\left(\frac{\rhos}{\rhon}\right)^{1/2}Q^{1/4}\nu^{-1/2},
	\label{eq:k2}
	\\[12pt]
	k_+ &=& \alpha^{3/2}Q^{3/4}\epsilon^{-1/2},
	\label{eq:k3}
\end{eqnarray}
where $\nu=\nu^n$ and $\epsilon\equiv-dE/dt$ is the mean energy dissipation (per unit mass), which, in the inertial subrange, is approximately constant. In the context of shell models we define
\begin{equation}
\epsilon_m=|u_m|^3k_m.
\end{equation}
We furthermore find, due to the efficient locking of the superfluid and normal fluid velocities in the inertial subrange, that $\epsilon^n\approx\epsilon^s$, which we refer to as $\epsilon$ in Eqs.~(\ref{eq:k1},\ref{eq:k3}).

\begin{figure}
	\centering
	\includegraphics[width=0.6\textwidth,angle=-90]{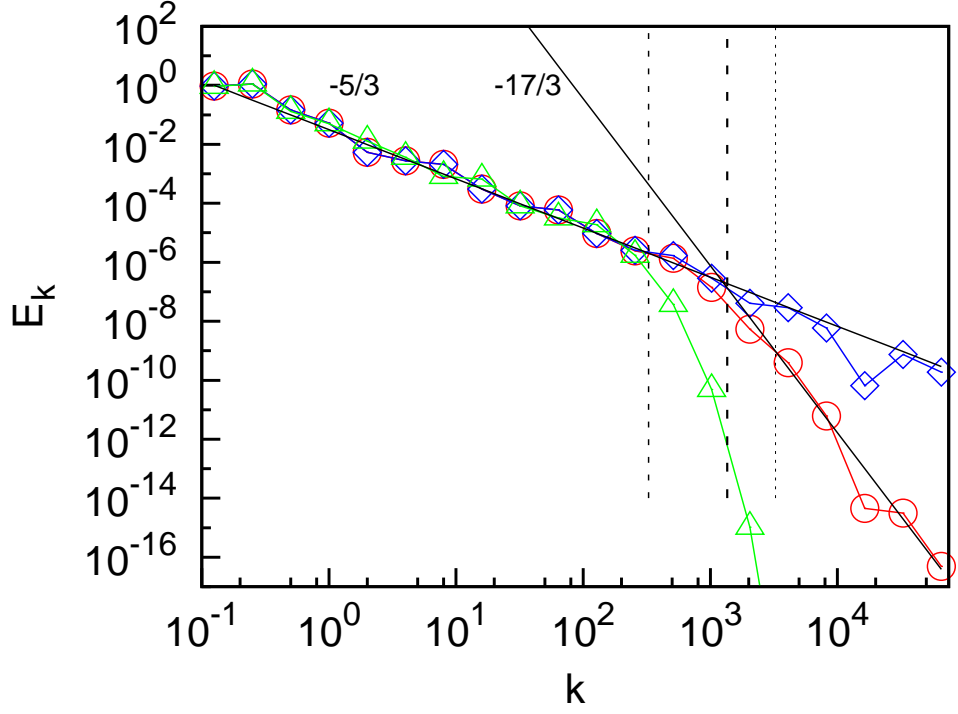}
	\caption{(Colour online). Log-log plot showing dissipation of turbulent energy in normal fluid at low temperature. Coupled normal fluid (red circles) is seen to deviate from uncoupled normal fluid (green triangles) as a result of mutual friction with superfluid (blue diamonds). Coupled normal fluid follows $k^{-5/3}$ for $k<k_{\eta}$, deviates slightly for $k_{\eta}<k<k_{\star}$ and follows a $k^{-17/3}$ power-law for $k>k_{\star}$. The vertical dotted line is $k_{\ell}$ and the short- and long-dashed lines are respectively $k_{\eta}$ and $k_{\star}$. For clarity $k_{\eta}<k_{\star}<k_{\ell}$.}
\label{fig4}
\end{figure}

We show the deviation at low temperature in (Fig.~\ref{fig4}), in which we show both power-laws and the critical wavenumbers at which the spectra change. We calculate these wavenumbers in accordance with the approximations derived by L'vov \etal \cite{lvov2006}. The deviations at high temperature are best realised by considering a system of turbulent superfluid $^3$He-B (see below).

\begin{figure}
	\centering
	\includegraphics[width=0.6\textwidth,angle=-90]{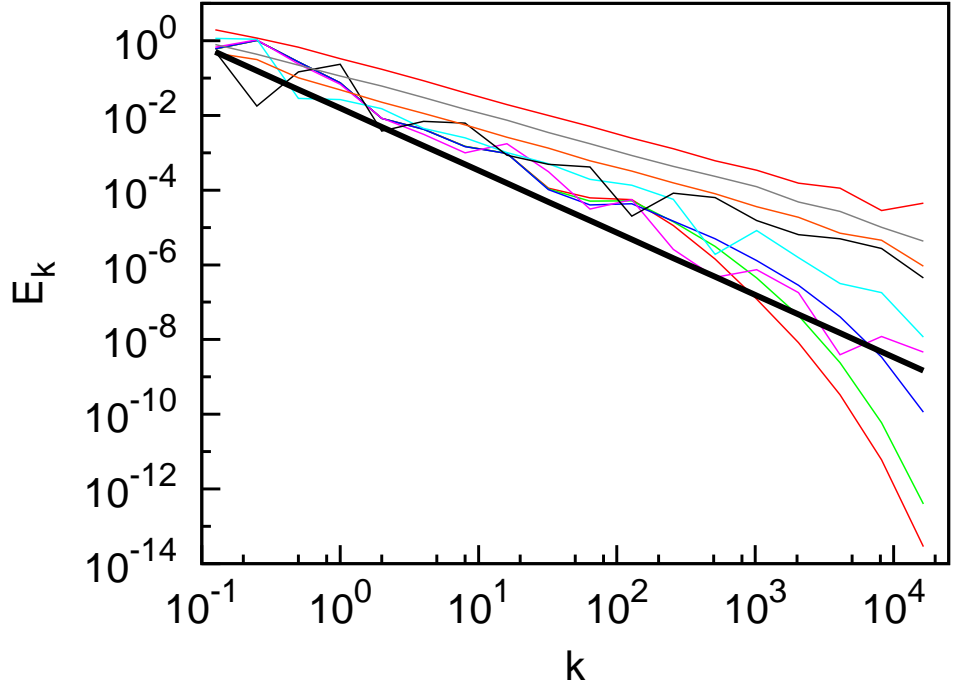}
	\caption{(Colour online). Log-log plot of development of build-up of energy in superfluid spectrum. Spectra (bottom to top) at time $t=0$ (red), $0.025$ (green), $0.05$ (dark blue), $0.5$ (pink), $5$ (light blue), $50$ (black), $500$ (orange), $5000$ (grey) and $10000$s (red) after lowering the temperature. $k_l$ (not shown) moves from $k\approx10^3$ to $k\approx2\cdot10^4$ during this period. The thick black line denotes the $k^{-5/3}$ Kolmogorov spectrum.}
\label{fig5}
\end{figure}

As mentioned above, we expect that at a sufficiently low temperature the superfluid energy spectrum builds up at high $k$. This is because the only energy sink for the superfluid is the mutual friction which depends on the presence of the normal fluid to dissipate the superfluid energy. We show the development of the build-up of energy in Fig.~\ref{fig5}. We produced Fig.~\ref{fig5} by allowing the fluids to attain fully-developed spectra at high temperature. We then reduced the temperature significantly to a finite, fixed temperature well below our previous low $T$ and monitored the changes in the superfluid spectrum over time. In order to maintain numerical stability, we fixed the values of $\rhos$, $\rhon$ and $\nun$. Thus the only temperature-dependent variable was $\alpha$, which alone dictates the magnitude of the mutual friction felt by the superfluid. The spectrum is seen to change as progressively more energy moves to higher $k$. This phenomenon highlights the necessity for an energy sink at high $k$ such as the Kelvin wave cascade and phonon emission.

In classical turbulence the Reynolds number, $Re=UD/\nu$, is a measure of the ratio of the inertial and viscous terms, where $U$ and $D$ are the large scale velocity and length scale respectively. The inertial range exists as long as $Re\gg~1$. In the same spirit we may define a quantum Reynolds number as the ratio of the inertial and general dissipative terms, be they due to viscous dissipation or mutual friction. Using this definition we may say that the Reynolds number for the superfluid is temperature-dependent, in agreement with \cite{Leveque}: both normal fluid and superfluid spectra extend to higher wavenumbers as the temperature is decreased.

\begin{figure}
	\centering
	\includegraphics[width=0.6\textwidth,angle=-90]{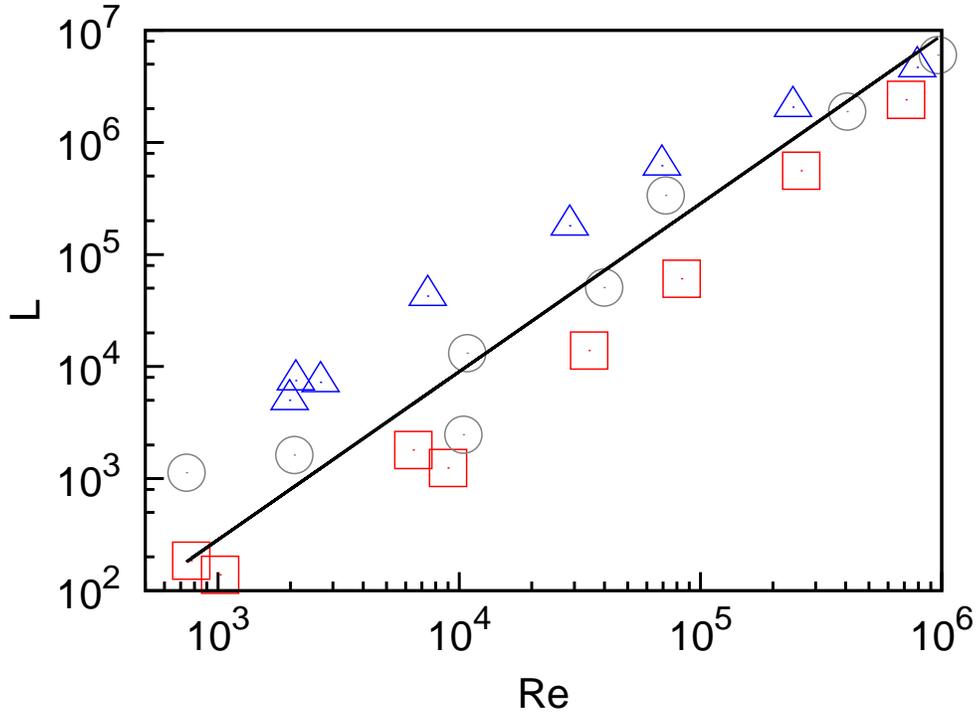}
	\caption{(Colour online).
    Log-log plot of vortex line density, $L$, vs Reynolds Number, $Re$, at all three temperatures: high (red squares), medium (grey circles) and low (blue triangles). The points from left to right correspond to increasing forcing, the leftmost point using forcing $f=(1+i)\cdot5\cdot10^{-6}$ and the rightmost point $f=(1+i)\cdot5\cdot10^1$. The Reynolds Number is calculated at the first shell ($n=1$) giving $D=2^4$. The solid black line is $L=Re^{3/2}$.}
\label{fig6}
\end{figure}

We furthermore show the dependence of the vortex line density, $L$, on the Reynolds number. According to Kolmogorov, the ratio of the dissipation scale and the scale of the large eddies is $\delta/D\propto~Re^{-3/4}$. In superfluid turbulence the smallest scale is $\ell\propto~L^{-1/2}$, hence we expect $L\propto~Re^{3/2}$. We show that this scaling holds true by considering $Re$ and $L$ and allowing the forcing to vary between $f=(1+i)\cdot5\cdot10^{-6}$ and $f=(1+i)\cdot5\cdot10^1$ in multiples of $10$ (see Fig.~\ref{fig6}). This finding agrees with the truncated DNS model of Salort \etal \cite{SRL}.

\subsection{Decaying turbulence}
\begin{figure}
	\centering
	\includegraphics[width=0.6\textwidth,angle=-90]{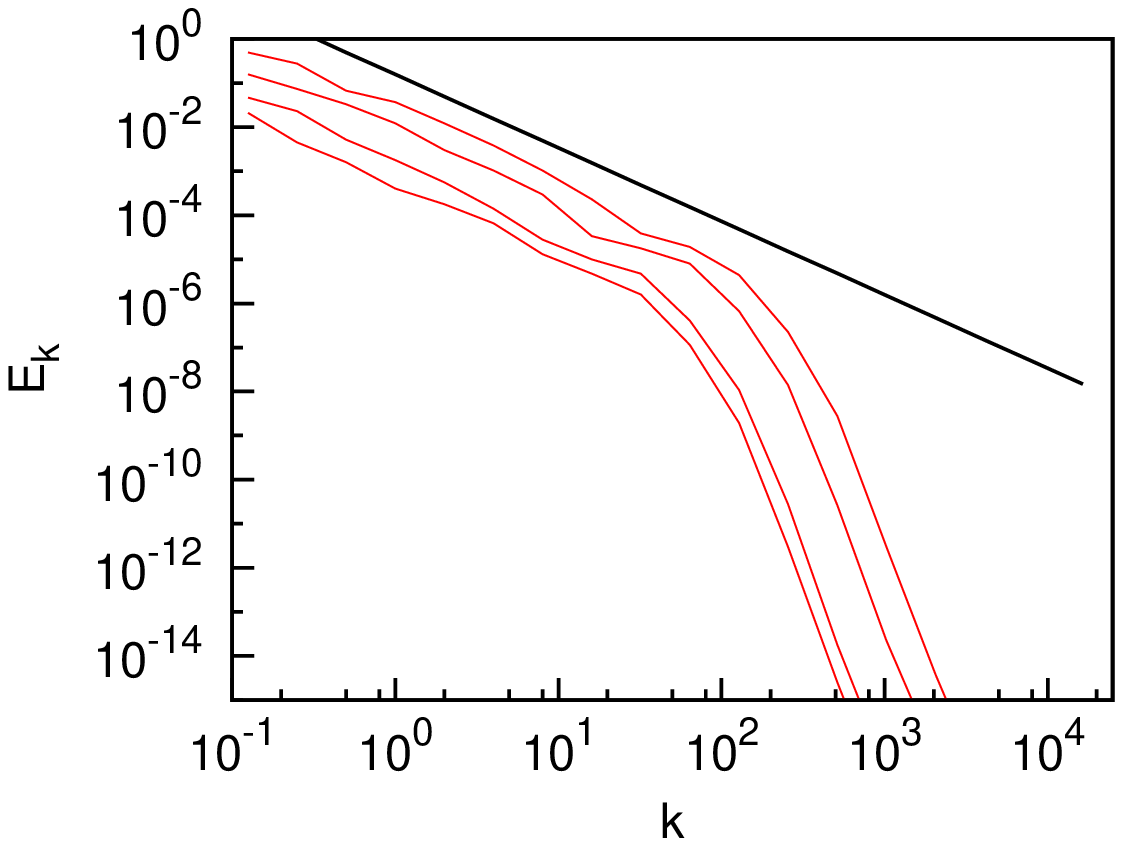}
	\includegraphics[width=0.6\textwidth,angle=-90]{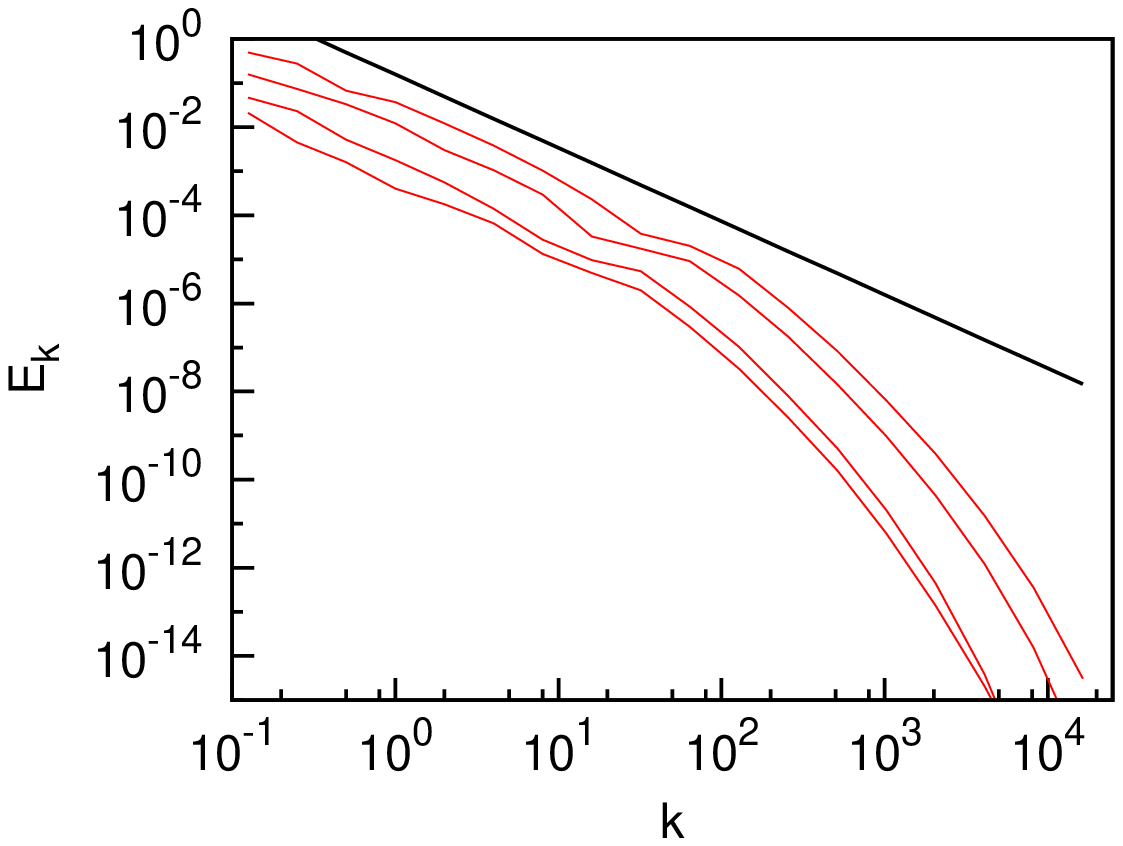}
	\caption{(Colour online). Decay of energy spectra for normal fluid (top) and superfluid (bottom) over $10$ realisations. Spectra top to bottom: After $500$, $1000$ , $2500$ and $5000$s. Also shown is the $k^{-5/3}$ spectrum (solid black line).}
\label{fig7}
\end{figure}

\begin{figure}
	\centering
	\includegraphics[width=0.6\textwidth,angle=-90]{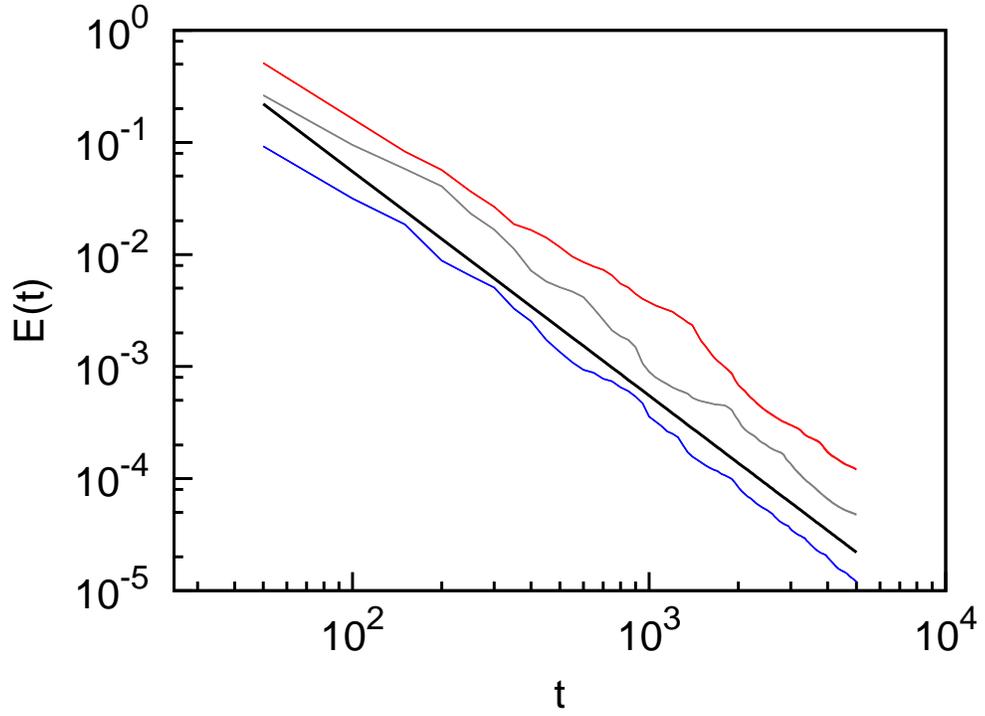}
	\caption{(Colour online). Decay of total energy over period of $5000$s ensemble averaged over $10$ realisations. Top to bottom: High (red line), medium (grey line) and low (blue line) temperatures. Shifted to show power law. Also shown is the $t^{-2}$ spectrum (solid black line).}
\label{fig8}
\end{figure}

\begin{figure}
	\centering
	\includegraphics[width=0.6\textwidth,angle=-90]{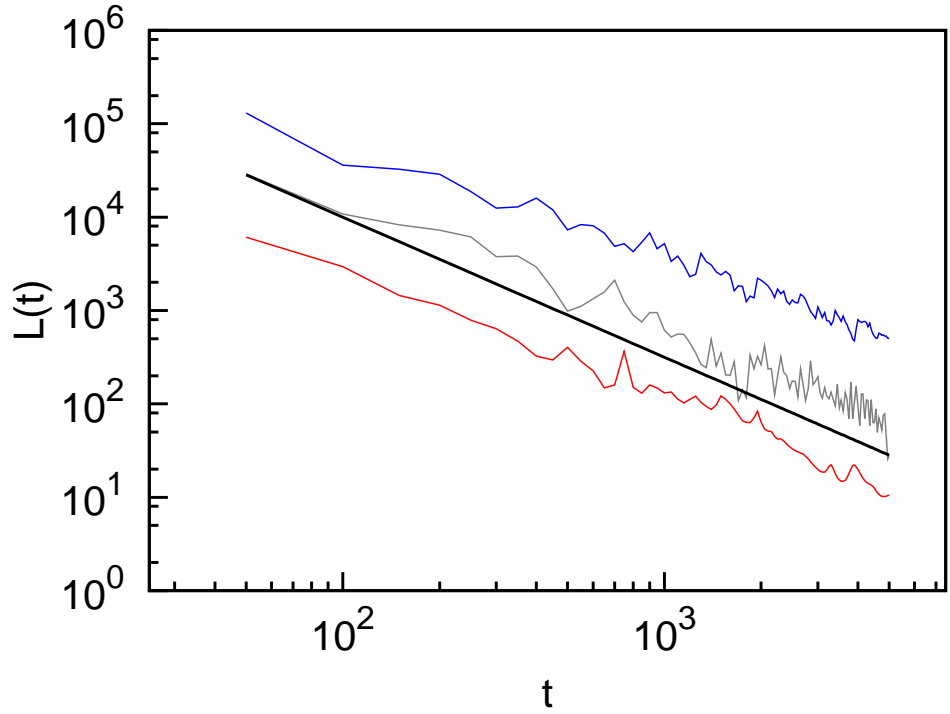}
	\caption{(Colour online). Decay of vortex line density over period of $5000$s ensemble averaged over $10$ realisations. Top to bottom: Low (blue line), medium (grey line) and high (red line) temperatures. Data is not shifted. Also shown is the $t^{-3/2}$ spectrum (solid black line).}
\label{fig9}
\end{figure}

To study the decay of turbulence we let $f=0$ and start from saturated spectra as an initial condition. We observe that during the decay both the superfluid and normal fluid spectra maintain their initial shape ($k^{-5/3}$ at low to intermediate values of $k$, followed by a more rapid drop at larger $k$), as shown in Fig.~\ref{fig7} (top and bottom). The spectra shown in these figures are the result of ensemble averaging over $10$ realisations. We also found that the total turbulent kinetic energy, $E(t)$, where
\begin{equation}
E^j(t)=\sum_m\frac{1}{2}|u_m^j(t)|^2,~(j=n,s),
\end{equation}
and the vortex line density, $L(t)$, decay as $E(t)\propto~t^{-2}$ and $L(t)\propto~t^{-3/2}$ as shown in Figs~\ref{fig8} and~\ref{fig9}, which is in agreement with experiments \cite{stalp,walm1,walm2} and theoretical models \cite{SND}.

\subsection{Forced turbulence in superfluid $^3$He-B}
The other non-radioactive isotope of liquid helium is $^3$He. $^3$He atoms are fermions, having only a single neutron in their nuclei, and achieve superfluidity by the Cooper pairing of two atoms. Although $T_c$ for $^4$He is about $2.17$~K, for $^3$He it is much lower, at about 1~mK. Experimental studies of $^3$He have focused on the so-called B-phase, known as $^3$He-B. The most notable physical property of $^3$He-B is the rapid increase in the normal fluid viscosity below $T_c$, which is proportional to $1/T^2$ \cite{landau}. In fact the viscosity is so great that for all intents and purposes the normal fluid may be considered to be at rest. We may thus consider such a system to be that of a turbulent superfluid in the presence of a stationary normal fluid. This description is equally applicable to that of high temperature $^4$He in the wavenumber subrange $k\gg~k_{\eta}$. This correspondence has already been noted by L'vov \etal \cite{lvov2006}. A second relevant physical property is the change in its quantum of circulation such that $\kappa=h/2m_3$, or about $2/3$ that of $^4$He, where $m_3$ denotes the mass of a $^3$He atom.

In a theoretical paper Vinen \cite{vinen3he} showed that in superfluid $^3$He-B the positions of the dissipative subrange and the inertial subrange are reversed, such that the dissipation due to the mutual friction occurs at low wavenumbers whereas the inertial subrange obeying the $k^{-5/3}$ power-law is located at high wavenumbers. His prediction for the wavenumber at which this changeover takes place is in agreement with that of L'vov \etal \cite{lvov2004,lvov2006}. L'vov \etal \cite{lvov2004} furthermore predicted that the dissipation should follow a $k^{-3}$ power-law, the same power law as that of high temperature $^4$He in the wavenumber subrange $k\gg~k_{\eta}$ \cite{lvov2006}.

\begin{figure}
	\centering
	\includegraphics[width=0.6\textwidth,angle=-90]{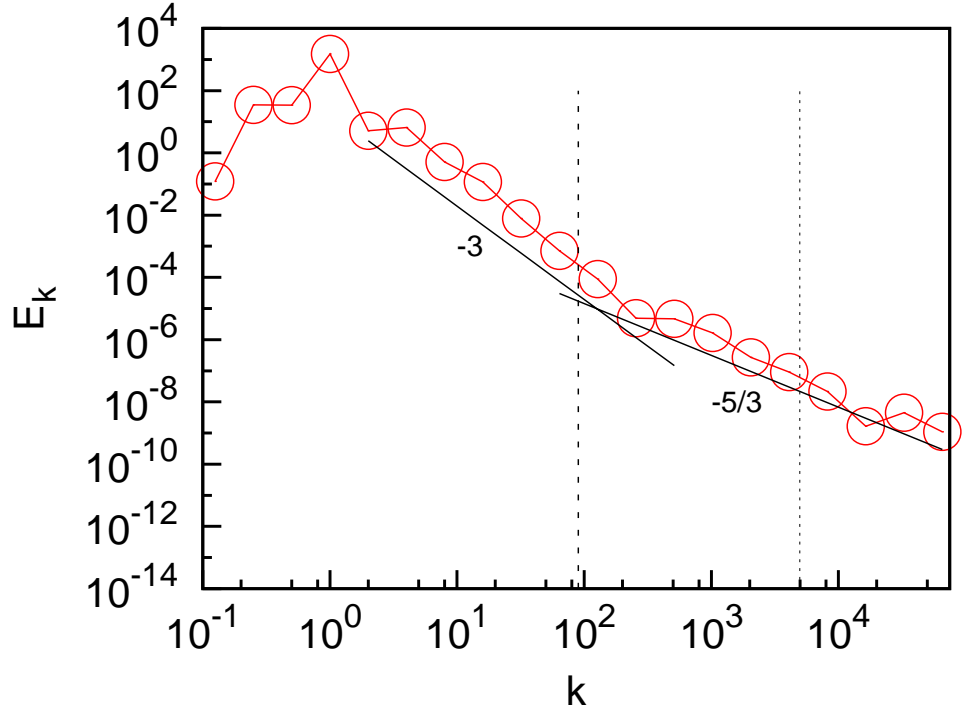}
	\caption{(Colour online). Log-log plot of fully developed, steady power spectrum for superfluid (red circles) in the presence of a stationary normal fluid in $^3$He-B. We show the two power-laws, $k^{-3}$ and $k^{-5/3}$. The dashed line is $k_+$, the wavenumber at which the inertial term becomes of the same order as the mutual friction term, and the dotted line is $k_{\ell}$, the intervortex spacing.}
\label{fig10}
\end{figure}

In order to model turbulence in $^3$He-B we reduced our two-fluids shell model back to a single-fluid model, replacing viscous dissipation with a dissipation due to mutual friction. The shell model equation then becomes
\begin{equation}
\frac{d}{dt} u_m^s 
=G_m[u^s] - F_m + f^s \delta_{m,m'},
\label{eq:goy-he3}
\end{equation}
where the mutual friction is now of the form
\begin{equation}
F_m=\alpha \kappa^{^3He} L u_m^s.
\end{equation}
The temperature dependence of Eq.~\ref{eq:goy-he3} is controlled by the parameter $\alpha$ alone. Experience has shown that a mutual friction of this form is extremely efficient at dissipating energy and in order to attain a steady spectrum of the kind described in \cite{vinen3he,lvov2004,lvov2006} it is necessary both to reduce $\alpha$ to $O(10^{-2})$ (much below $T_c$ and very close to a pure superfluid state) and to increase the forcing to $O(10^2)$ ($5$ orders of magnitude greater than that used for $^4$He). The numerical results show the spectrum initially decreasing like $k^{-3}$ which subsequently changes to $k^{-5/3}$. The changeover wavenumber is in reasonably good agreement with $k_+$ predicted by L'vov (see Fig.~\ref{fig10}). We used the following parameters $(M,\lambda,m')=(20,2,4)$.

\section{Discussion}
We have developed a two-fluids shell model based on the GOY shell model of classical turbulence. We have shown that, in continually excited turbulence, the mutual friction effectively locks together the normal fluid and superfluid over a wide range of wavenumbers, forming a $k^{-5/3}$ Kolmogorov spectrum.
By reducing the temperature and hence the normal fluid fraction, we have shown that this range's cut-offs are temperature-dependent for both the normal fluid and superfluid.
In other words, lowering the temperature leads to longer inertial subranges.
This result for the energy spectrum agrees with the experiment of Tabeling \cite{Tabeling} and the DNS of Roche \etal \cite{Leveque}.
The relation which we find between vortex line density and Reynolds number agrees with Salort \etal \cite{SRL}.
We have realised the deviations from the $k^{-5/3}$ spectrum due to the mutual friction predicted by L'vov \cite{lvov2006} for $^4$He at low temperature and by Vinen \cite{vinen3he} and L'vov \cite{lvov2004,lvov2006} for $^3He$-B. 
In the zero-temperature limit our model develops a build-up of energy at high wavenumbers, highlighting the necessity for an energy sink at high $k$, such as the Kelvin wave cascade and phonon emission \cite{Bag11b}.
Our model shares the common advantage of all shell models: being flexible and computationally inexpensive, it is relatively easy to investigate the decay of turbulence, which would be more difficult with DNS.
Our results for the decay of turbulence are in agreement with experiments \cite{stalp,walm1,walm2} and theoretical models \cite{SND}.
We anticipate that our two-fluids model can be used to study other aspects of quantum turbulence.\bigskip

ACKNOWLEDGEMENTS

One of the authors (CFB) is grateful to F. Toschi for stimulating interest in shell models. We also thank V.S L'vov for his comments. This work was made possible with support from the EPSRC.


\end{document}